\documentclass[10pt]{iopart}
\usepackage{graphicx}
\def\lesssim{\mathrel{\hbox{\rlap{\hbox{\lower4pt\hbox{$\sim$}}}\hbox{$<$}}}}
\def\gtrsim{\mathrel{\hbox{\rlap{\hbox{\lower4pt\hbox{$\sim$}}}\hbox{$>$}}}}
\def\MNRAS{{\it Mon.\ Not.\ Royal Astron.\ Soc. }}
\def\MNRASL{{\it Mon.\ Not.\ Royal Astron.\ Soc.\ Lett. }}
\def\ApJ{{\it Astroph.\ J. }}
\def\ApJL{{\it Astroph.\ J.\ Lett. }}
\def\ApJS{{\it Astroph.\ J.\ Supp. }}
\def\AJ{{\it Astron.\ J. }}
\def\PRL{{\it Phys.\ Rev.\ Lett. }}
\def\PRD{{\it Phys.\ Rev.\ }D }

\begin{document}

\title[EM Counterparts to BH Mergers]{Electromagnetic Counterparts to
  Black Hole Mergers}
\author{Jeremy D.\ Schnittman$^{1,2}$}
\address{$^1$ Department of Physics and Astronomy, Johns Hopkins
  University, Baltimore, MD 21218} 
\address{$^2$ NASA Goddard Space Flight Center, Greenbelt, MD 20771}

\begin{abstract}
During the final moments of a binary black hole (BH) merger, the
gravitational wave (GW) luminosity of the system is greater than the
combined electromagnetic output of the entire observable
universe. However, the extremely weak coupling between GWs and
ordinary matter
makes these waves very difficult to detect directly. Fortunately,
the inspiraling BH system will interact strongly--on a purely
Newtonian level--with any surrounding material in the host galaxy, and
this matter can in turn produce unique electromagnetic (EM) signals
detectable at Earth. By identifying EM counterparts to GW sources,
we will be able to study the host environments of the merging BHs, in
turn greatly expanding the scientific yield of a mission like LISA. 
\end{abstract}

\pacs{95.30.Sf, 98.54.Cm, 98.62.Js, 04.30.Tv, 04.80.Nn}
\submitto{\CQG}
\maketitle

\section{INTRODUCTION}
\label{intro}

Prompted by recent advances in numerical relativity (NR), there has
been an increased
interest in the astrophysical implications of black hole (BH)
mergers (see \cite{decadal} for a sample of related White Papers submitted
to the recent Astro2010 Decadal Report). Of particular interest is the
possibility of a distinct,
luminous electromagnetic (EM) counterpart to a gravitational-wave (GW)
signal. If such an EM counterpart could be identified
with a LISA$^{\footnote{\tt http://lisa.nasa.gov}}$
detection of a supermassive BH binary in the merging process, then the
host galaxy could likely be determined
\cite{kocsis:06,lang:06,lang:08,kocsis:08a}. Like the cosmological
beacons
of gamma-ray bursts and quasars, merging BHs can teach us about
relativity, high-energy astrophysics, radiation hydrodynamics, dark
energy, galaxy formation and evolution, and even dark matter.
A large variety of potential
EM signatures have recently been proposed, almost all of which require
some significant amount of gas in the near vicinity of the merging
BHs. In this paper, we review the recent literature on EM signatures,
and propose a rough outline of the future work, both observational and
theoretical, that will be needed to fully realize the potential of GW
astronomy.

\section{DIVERSITY OF SOURCES}
\label{sources}

\begin{figure}
\begin{center}
\includegraphics[width=\textwidth]{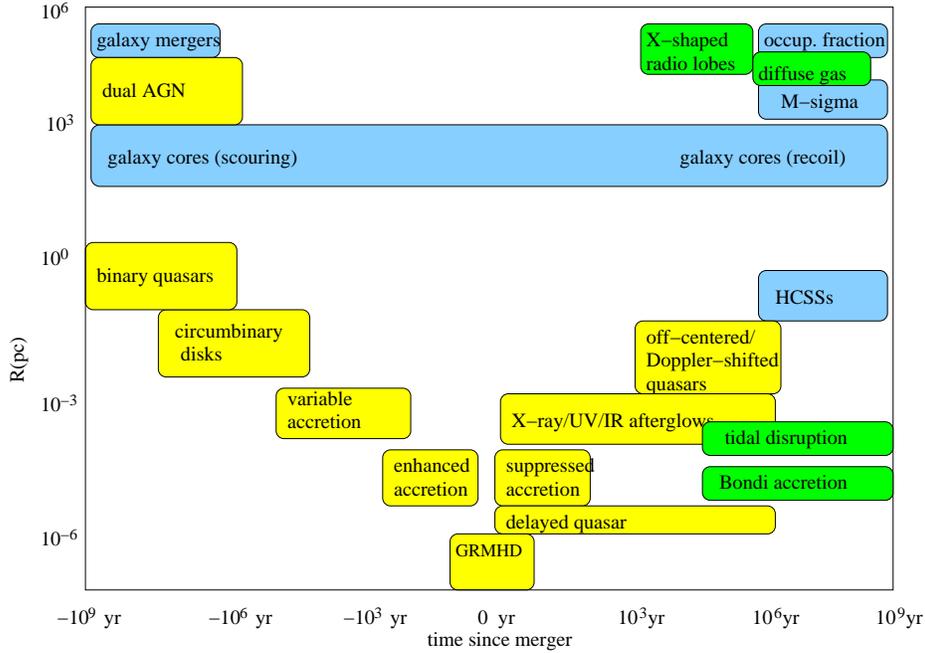}
\caption{\label{source_chart} Selection of potential EM signatures,
  sorted by timescale, typical size of emission region, and physical
  mechanism (blue = stellar; yellow = accretion disk; green = diffuse
  gas/miscellaneous).}
\end{center}
\end{figure}

From a theoretical point of view, EM signatures can be 
categorized by the physical mechanism responsible for the emission,
namely stars, hot diffuse gas, or circumbinary/accretion disks. In
Figure \ref{source_chart}, we show the diversity of these sources,
arranged according the spatial and time scales on which they occur.

It is important to note that, while the black holes themselves are of
course extremely relativistic objects, most of the observable effects
occur on distance and time scales that are solidly in the Newtonian
regime. While one of the most interesting NR results in recent years
has been the prediction of large recoil velocities originating from
the final merger and ringdown of binary BHs \cite{bigkicks}, the {\it
  astrophysical} implications of these large kicks are for the most
part entirely Newtonian. 

\subsection{Stellar Signatures}

On the largest scales, we have strong circumstantial evidence of
supermassive BH mergers at the centers of merging galaxies. From large
optical surveys of interacting galaxies out to redshifts of $z \sim
1$, we can infer that $5-10\%$ of massive galaxies
are merging at any given time, and the majority of galaxies with
$M_{\rm gal} \gtrsim 10^{10} M_\odot$ have experienced a major merger
in the past 3 Gyr \cite{bell:06,mcintosh:08,deravel:09,bridge:10},
with even higher merger
rates at redshifts $z\sim 1-3$ \cite{conselice:03}. At the same time,
high-resolution observations of nearby galactic nuclei find that every
large galaxy hosts a SMBH in its center \cite{kormendy:95}. Yet we see
a remarkably small number of dual AGN \cite{komossa:03,comerford:09},
and only one known source with an actual binary system where the BHs
are gravitationally bound to each other \cite{rodriguez:06}. Taken
together, these observations strongly suggest that when galaxies
merge, the merger of their central SMBHs inevitably follows, and
likely occurs on a relatively short time scale, which would explain
the apparent scarcity of binary BHs. 

There is also indirect evidence for SMBH mergers in the stellar
distributions of galactic nuclei, with many elliptical galaxies
showing light deficits (cores), which correlate strongly with the
central BH mass \cite{kormendy:09}. The cores are evidence of a
history of binary BHs that scour out the nuclear stars via three-body
scattering
\cite{milosavljevic:01,milosavljevic:02,merritt:07}, or even
post-merger relaxation of recoiling BHs
\cite{merritt:04,boylan-kolchin:04,gualandris:08,guedes:09}. 

While essentially all massive nearby galaxies appear to
host central SMBHs, it is quite possible that this is not the case
at larger redshifts and smaller masses, where major mergers could lead
to the complete ejection of the final black hole via large
gravitational-wave recoils. By measuring the occupation fraction of
BHs in distant galaxies, one could infer merger rates and the
distribution of kick velocities \cite{schnittman:07a,volonteri:07,
  schnittman:07b,volonteri:08a,volonteri:10}. The occupation fraction
will of course also affect the LISA event rates, especially at high
redshift \cite{sesana:07}. An indirect signature for kicked BHs
could potentially show up in the statistical properties of active
galaxies, in particular in the relative distribution of different
classes of AGN in the ``unified model'' paradigm \cite{komossa:08b,blecha:10}.
On a smaller scale, the presence of
intermediate-mass BHs in globular clusters also gives indirect
evidence of their merger history \cite{holley-bockelmann:08}.

Another EM signature of BH mergers comes from the population of
stars that remain bound to a recoiling black hole that gets ejected
from a galactic nucleus \cite{komossa:08a,merritt:09,oleary:09}. These
stellar systems will appear similar to
globular clusters, yet with smaller spatial extent and much larger
velocity dispersions, as the potential is completely dominated by the
central SMBH. 

\subsection{Gas Signatures: Accretion Disks}

Gas in the form of accretion disks around single massive BHs is known to
produce some of the most luminous objects in the universe. However,
very little is known about the behavior of accretion disks around {\it
  two} BHs, particularly at late times in their inspiral evolution. In
Newtonian systems, it is believed that a circumbinary accretion disk
will have a central gap of much lower density, either
preventing accretion altogether, or at least decreasing it
significantly \cite{pringle:91, artymowicz:94, artymowicz:96}. When
including the evolution of the binary due
to GW losses, the BHs may also decouple from the disk at the point
when the GW inspiral time becomes shorter than the gaseous inflow time at
the inner edge of the disk \cite{milos:05}. This decoupling should
effectively stop accretion onto the central object until the gap can
be filled on an inflow timescale. However, other semi-analytic
calculations predict an {\it enhancement} of accretion power as the evolving
binary squeezes the gas around the primary BH, leading to a rapid
increase in luminosity shortly before merger
\cite{armitage:02,chang:10}. 

Regardless of {\it how} the gas can or
cannot reach the central BH region, a number of recent papers have
shown that if there {\it is} sufficient gas present, then an observable
EM signal is likely. Krolik \cite{krolik:10} used analytic arguments to
estimate a peak luminosity comparable to that of the Eddington limit,
independent of the detailed mechanisms for shocking and heating the
gas. Using relativistic magneto-hydrodynamic simulations in 2D,
O'Neill \etal
\cite{oneill:09} showed that the prompt mass loss due to GWs may actually
lead to a sudden {\it decrease} in luminosity following the merger, as the
gas in the inner disk temporarily has too much energy and angular momentum to
accrete efficiently. Full NR simulations of the final few orbits of a
merging BH binary have now been carried out including the presence of
EM fields in a vacuum \cite{palenzuela:09, mosta:10, palenzuela:10}
and also gas, treated as test particles in \cite{vanmeter:10} and as
an ideal fluid in \cite{bode:10} and \cite{farris:10}. The
simulations including matter all suggest that the gas can get shocked
and heated to high temperatures, thus leading to bright counterparts
in the event that sufficient gas is in fact present in the immediate
vicinity of the merging BHs.

If the primary energy source for heating the gas is gravitational,
then typical efficiencies will be on the order of $\sim 1-10$\%,
comparable to that expected for standard accretion in
AGN. However,
if the merging BH binary is able to generate strong magnetic
fields \cite{palenzuela:09, mosta:10, palenzuela:10}, then highly
relativistic jets may be launched along the resulting BH spin axis,
converting matter to energy with a Lorentz boost factor of $\Gamma \gg 1$. 
Even with purely hydrodynamic heating,
particularly bright and long-lasting afterglows may be produced
in the case of very large recoil velocities, which effectively can
disrupt the entire disk, leading to strong shocks and dissipation
\cite{lippai:08, shields:08, schnittman:08, megevand:09, rossi:10,
  anderson:10, corrales:10, tanaka:10a, zanotti:10}. For systems that open up a
gap in the circumbinary disk, an EM signature may take the form of a
quasar suddenly turning on as the gas refills the gap, months to years
after the BH merger \cite{milos:05, shapiro:10, tanaka:10b}. 

For those systems that also received a large kick at the time of
merger, we may observe quasar activity for millions of years after,
with the source displaced from the galactic center, either spatially
\cite{kapoor:76, loeb:07,
  volonteri:08b,civano:10,dottori:10,jonker:10} or spectroscopically
\cite{bonning:07, komossa:08c,boroson:09,robinson:10}. However, large offsets between the
redshifts of quasar emission lines and their host galaxies have also
been interpreted as evidence of pre-merger binary BHs
\cite{bogdanovic:09,dotti:09,tang:09,dotti:10b} or due to the large relative
velocities in merging galaxies \cite{heckman:09,shields:09a,vivek:09,decarli:10}, or
``simply'' extreme examples of the class of double-peaked emitters,
where the line offsets are generally attributed to the disk
\cite{gaskell:88,eracleous:97,shields:09b,chornock:10,gaskell:10}. 

In addition to the many potential prompt and afterglow signals from
merging BHs, there has also been a significant amount of
theoretical and observational work focusing on the early precursors of
mergers. Following the evolutionary trail from the upper-left of
Figure 1, we see that shortly after a galaxy merges, dual AGN may
form with typical separations of a few kpc
\cite{komossa:03,comerford:09}, sinking to the center of the merged
galaxy on a relatively short timescale ($\lesssim$ 1 Gyr) due to
dynamical friction \cite{begelman:80}. This merger process is
also expected to funnel a great deal of gas to the galactic center, in
turn triggering quasar activity
\cite{hernquist:89,kauffmann:00,hopkins:08,green:10}. At separations
of $\sim 1$ pc, the BH binary (now ``hardened'' into a gravitationally
bound system) could stall, having depleted its loss cone of stellar
scattering and not yet reached the point of gravitational radiation
losses \cite{milosavljevic:03}. Gas dynamical drag from
massive disks ($M_{\rm disk} \gg M_{\rm BH}$) leads to a prompt
inspiral ($\sim 1-10$ Myr), in most cases able to reach sub-parsec
separations, depending on the resolution of the simulation
\cite{escala:04,kazantzidis:05,escala:05,dotti:07,
cuadra:09,dotti:09b,dotti:10a}.

At this point, a proper binary quasar is formed, with an orbital period
of months to decades, which could be identified by periodic accretion
\cite{macfadyen:08, hayasaki:08, haiman:09a, haiman:09b} or red-shifted broad emission
lines as mentioned above \cite{bogdanovic:08,shen:09,loeb:10}. Direct
GW stresses on the circumbinary disk might also lead to periodic
variations in the light curve, although with very small amplitude
\cite{kocsis:08}. 

\subsection{Gas Signatures: Diffuse Gas; ``Other''}
In addition to the many disk-related signatures, there are also a
number of potential EM counterparts that are caused by the accretion
of diffuse gas in the galaxy. For BHs that get significant kicks at
the time of merger, we expect to see quasi-periodic episodes of Bondi
accretion as the BH oscillates through the gravitational potential
of the galaxy over millions of years, as well as off-center AGN
activity \cite{blecha:08, fujita:09,guedes:10,sijacki:10}. On larger
spatial scales, the recoiling BH could also produce trails of
overdensity in the hot interstellar gas of elliptical galaxies
\cite{devecchi:09}. In a similar way, rogue SMBHs in gas-rich galaxies
could leave trails of star formation in their wake
\cite{fuente:08}. It is even possible that the same density
enhancements could be detected via off-nucleus gamma-ray emission from
annihilating dark matter particles \cite{mohayaee:08}.
Also on kpc--Mpc scales, X-shaped radio jets have been seen in a
number of galaxies, which could possibly be due to the merger and
subsequent spin-flip of the central BHs \cite{merritt:02}.

Another potential source of EM counterparts comes not from diffuse
gas, or accretion disks, but the occasional capture and tidal
disruption of normal stars by the merging BHs. This tidal disruption,
which also occurs in ``normal'' galaxies \cite{rees:88,komossa:99,halpern:04},
may be particularly easy to identify in off-center BHs following a
large recoil \cite{komossa:08a}. Tidal disruption rates may be strongly
increased by the merger process itself
\cite{chen:09,stone:10,seto:10,schnittman:10}, while the actual
disruption signal may be truncated by the pre-merger binary
\cite{liu:09}. These events are likely to be seen by the dozen in
coming years with PanSTARRS and LSST \cite{gezari:09}. In addition to the tidal disruption scenario, in
\cite{schnittman:10} we showed how gas or stars trapped at the stable
Lagrange points in a BH binary could evolve during inspiral and
eventually lead to enhanced star formation, ejected hyper-velocity
stars, highly-shifted narrow emission lines, and short bursts of
Eddington-level accretion coincident with the BH merger. 

A completely different type of EM counterpart can be seen in the
radio. Namely, nanosecond time delays in the arrival of pulses from
millisecond radio pulsars is direct evidence of extremely
low-frequency (nano-Hertz) gravitational waves from massive ($\gtrsim
10^8 M_\odot$) BH binaries
\cite{jenet:06,sesana:08,sesana:09,jenet:09,seto:09,pshirkov:10,vanhaasteren:10,sesana:10}.
By cross-correlating the signals from multiple pulsars around the sky,
we can effectively make use of a GW detector the size of the entire
galaxy. 

\section{GAME PLAN}
In the coming years, a number of theoretical and observational
advances will be required in order to fully realize the potential of
GW/EM multi-messenger astronomy. Some of the central questions that
need to be answered include:
\begin{itemize}
\item What is the galaxy merger rate as a function of galaxy mass,
  mass ratio, gas fraction, cluster environment, and redshift?
\item What is the mass function and spin distribution of the central
  BHs in these merging (and non-merging) galaxies?
\item What is the central environment around the BHs, prior to merger?
  \begin{itemize}
  \item What is the quantity and quality (temperature, density,
    composition) of gas?
  \item What is the stellar distribution (age, mass function,
    metallicity)?
  \item What are the properties of the circumbinary disk?
  \end{itemize}
\item What is the time delay between galaxy merger and BH merger?
\end{itemize}
We have rough predictions for some of these questions from
cosmological N-body simulations, but the uncertainties and model
dependencies are quite large. Similarly, observational constraints are
currently quite weak and often open to widely varying interpretations.

\subsection{Theory}
With respect to the questions outlined above, improved cosmological
simulations will certainly help improve our estimates for galactic and
BH merger rates, as well as the gas environments expected in the
central regions. Particularly promising are multi-scale simulations
that can zoom in on regions of interest, going to higher resolution
and more realistic physics closer to the BHs \cite{springel:05}. To
model more accurately the interaction between the circumbinary disk
and the BHs, grid-based methods (as opposed to smoothed particle
hydrodynamics; SPH) will be necessary, especially at the inner edge
where steep density and pressure gradients are likely to be found. The
accurate treatment of this region is critical to understand the gas
environment immediately around the BHs at time of merger, and thus
whether any bright EM signal is likely to be produced. 

The natural product of these (Newtonian) circumbinary MHD simulations
would be a set of reasonable initial conditions to be fed into the
much more computationally intensive NR codes that compute the final
orbits and merger of the BHs, now including matter and magnetic
fields. The results of
\cite{palenzuela:09,mosta:10,palenzuela:10,vanmeter:10,bode:10,farris:10}
are extremely impressive from a computational point of view, but their
astrophysical relevance is limited by our complete ignorance of
the likely initial conditions. Even with perfect knowledge of the
initial conditions, the value of the MHD simulations is also limited
by the lack of radiation transport and accurate thermodynamics, which
are only now being incorporated into local Newtonian simulations of
steady-state accretion disks \cite{hirose:09}. Significant future work
will be required to incorporate the radiation transport into a fully
relativistic global framework, required not just for accurate modeling
of the dynamics, but also for the prediction of EM signatures that
might be compared directly with observations.

\subsection{Observations}

Even with the launch of LISA a decade or more away, many of the EM
counterparts discussed above should be observable today, in some cases
even giving unambiguous evidence for merging BHs. On the largest
distance and time scales, dual AGN candidates can be identified with
large spectroscopic surveys like SDSS$^{\footnote{\tt
  http://www.sdss.org}}$, then followed up with high-resolution imaging
and spectroscopy. Combined with surveys of galaxy morphology and
pairs, the distribution of dual AGN will help us test theories of
galactic merger rates as a function of mass and redshift, as well as
the connection between gas-rich mergers and AGN activity. 
Spectroscopic surveys should also be able to identify many candidate
binary AGN, which may be confirmed or ruled out with subsequent
observations over relatively short timescales ($\sim 1-10$ yrs), as
the line-of-site velocities to the BHs changes by an observable
degree. Long-lived afterglows could be discovered in existing
multi-wavelength surveys, but successfully identifying them as merger
remnants as opposed to obscured AGN or other bright unresolved sources
would require improved pipeline analysis of literally millions of
point sources, as well as extensive follow-up observations. 

Particularly promising as unambiguous examples of recoiling BHs would
be the measurement of large velocity dispersions in nearby ($d
\lesssim 20$ Mpc) globular clusters \cite{merritt:09}. With
multi-object spectrometers
on large ground-based telescopes, this is also technically realistic
in the immediate future. Perhaps the most exciting direction for the
coming decade of astronomy is in the time domain. Optical telescopes
like PTF and PanSTARRS are already taking data from huge areas of the
sky with daily and even hourly frequency. These time-domain surveys
are ideally suited for looking for variability from binary BH systems
as precursors to merger. Especially promising would be the detection
of long-period variable AGN, ideally suited to extensive
multi-wavelength follow-up observations.

\end{document}